\documentclass[preprint,aps,amsmath]{revtex4}

\usepackage{graphicx}
\usepackage{bm}
\usepackage{graphicx}
\usepackage{amssymb,amsmath}
\usepackage{subfigure}

\begin{document}
\preprint{}

\title{On the effects of channel tortuosity on the close electromagnetic fields associated with lightning return strokes}

\author{J. Peer and A. Kendl}

\affiliation{Institute for Ion Physics and Applied Physics, University of
  Innsbruck, Technikerstr. 25, A-6020 Innsbruck, Austria}

\begin{abstract}

The electromagnetic fields associated with tortuous lightning channels are
usually characterised by a pronounced fine structure. This work investigates
and quantifies the effects of channel tortuosity on the return stroke
electromagnetic field shapes in the close lightning environment (the range up
to 100\,m from the lightning striking point). General equations for lightning
return stroke electromagnetic fields for arbitrarily located observation
points are derived from Maxwell's equations. In order to include arbitrary
channel shapes, the channel is described by a parametric representation in
Cartesian coordinates with the channel length as the free parameter. The
return stroke current required for the evaluation of the derived equations is
calculated from a current generation type model. The field computations show
that amplitudes and waveforms of the electromagnetic fields in the close
lightning environment are considerably influenced by the channel shape. In
particular, the induction component of the electric fields radiated by a
tortuous lightning channel differs significantly from that associated with a
straight channel. 

\end{abstract}

\maketitle

\section{Introduction}
\label{introduction}

In calculating the electromagnetic fields radiated by ground striking
lightning discharges, the lightning channel is usually idealised by a straight
and vertical line between cloud and ground. Due to the symmetry in such
assumptions, the equations for the electric and magnetic fields can be
formulated in cylindrical or spherical coordinates
\citep[e.g.][]{thottappillil1997,thottappillil1998,thottappillil2001,thottappillil2004,thottappillil2007}. 

The close electromagnetic fields associated with cloud-to-ground lightning,
however, is characterised by a pronounced fine structure which is due to the
tortuosity of the lightning channel. Recently, various approaches for the
computation of electric and magnetic fields radiated by tortuous lightning
channels have been published \citep[e.g.][]{lupo2000,song2007,zhao2009}. The
loss of radial symmetry in considering arbitrarily shaped channels
necessitates the introduction of Cartesian coordinates which implicates an
increase of complexity in deriving and solving the equations for the
electromagnetic fields \citep{zhao2009}. 
 
Section \ref{fieldequations} of our work presents a detailed derivation of
very general expressions for lightning return stroke fields. The equations are
formulated in Cartesian coordinates and include both arbitrarily shaped
lightning channels and arbitrarily located observation points. Moreover, a
height variable return stroke velocity is taken into account. Arbitrary
channel shapes are considered by the introduction of a parameter
representation for the lightning channel with the channel length as the free
parameter. Such a description allows the direct application of return stroke
models which predict the current distribution as a function of the length
coordinate of the lightning channel. 

In our work, the return stroke current distribution required for the field
computations is calculated on the basis of the current generation type model
proposed by Cooray et al.~\cite{cooray2004}. The model includes the attachment process and
enables thus to include the upward growing connecting leader. Moreover, it
allows to take into account a height variable return stroke velocity. The
model of  Cooray et al.~\cite{cooray2004} and the modifications done in our work are briefly
discussed in section \ref{application}. 

The equations derived in section \ref{fieldequations} are evaluated for both a
straight and a tortuous lightning channel. The tortuous channel used in the
computations is formed from randomly generated parameters. The effects of
channel tortuosity on the close electromagnetic fields are investigated in
section \ref{application}, too.  

The presented computations were done within the framework of an investigation
on the action of lightning electromagnetic pulses on biological
tissue. Biological tissue is mainly influenced by the electric fields which
are induced by the transient magnetic pulses associated with return
strokes. For this reason, a short paragraph of section \ref{application} deals
with the induced electric fields.

\section{Derivation of general equations for lightning return stroke electromagnetic fields}
\label{fieldequations}

\subsection{Geometry}
\label{geometry}

\begin{figure}[hbt]
   \includegraphics[height=9.0cm]{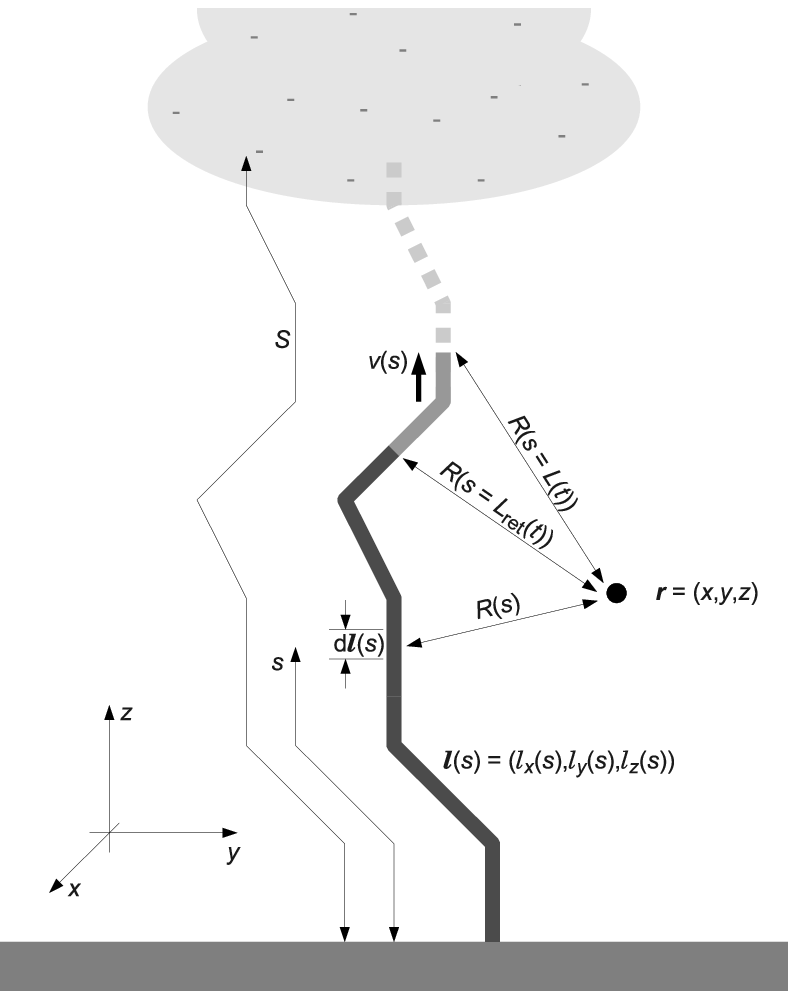}
   \caption{\it Geometry used in deriving equations for the electromagnetic fields
     radiated by a lightning return stroke. Real and retarded length of the
     lightning channel are illustrated by different shades of grey. The
     observation point is marked with $\vec{r}$. The grey bar indicates a
     perfectly conducting ground plane.}   
   \label{fig1}
\end{figure}

The geometry used in deriving equations for the electric and magnetic fields
radiated by a lightning return stroke is shown in figure \ref{fig1}. Cartesian
coordinates with basis vectors
$(\vec{\mathrm{e}}_x,\vec{\mathrm{e}}_y,\vec{\mathrm{e}}_z)$ are used. The
lightning channel
$\vec{l}(s)=l_x(s)\,\vec{\mathrm{e}}_x+l_y(s)\,\vec{\mathrm{e}}_y+l_z(s)\,\vec{\mathrm{e}}_z$
is parametrised by its length $s$, where $s$ is measured from ground on
upward. The base point of the channel is assumed to be the point of origin of
the coordinate
system. $R(s)=\sqrt{(\mathrm{d}\vec{l}(s)-\vec{r})\cdot(\mathrm{d}\vec{l}(s)-\vec{r})}$
denotes the distance between a channel segment $\mathrm{d}\vec{l}(s)$ and the
observation point
$\vec{r}=x\,\vec{\mathrm{e}}_x+y\,\vec{\mathrm{e}}_y+z\,\vec{\mathrm{e}}_z$. The
velocity distribution of the upward extending return stroke channel is assumed
to be specified by a function $v(s)$. Real and retarded channel lengths at
time $t$ are denoted by $s=L(t)$ and $s=L_\mathrm{ret}(t)$, respectively. 

\subsection{Solution of Maxwell's equations for the current carrying return stroke channel}

Maxwell's equations in vacuum are given by (since $\varepsilon_\mathrm{r}\approx 1$, $\mu_\mathrm{r}\approx1$ in air, air can be idealised as vacuum)
\begin{subequations}
   \label{eqmaxwell}
   \begin{align}
      &\vec{\nabla}\cdot\vec{E}(\vec{r},t)  = \frac{\rho(\vec{r},t)}{\varepsilon_0}\\
      &\vec{\nabla}\cdot\vec{B}(\vec{r},t)  = 0\phantom{\frac{1}{1}}\\
      \label{eqinduction}
      &\vec{\nabla}\times\vec{E}(\vec{r},t) = -\frac{\partial\vec{B}(\vec{r},t)}{\partial t}\\
      &\vec{\nabla}\times\vec{B}(\vec{r},t) =
      \mu_0\varepsilon_0\,\frac{\partial \vec{E}(\vec{r},t)}{\partial t} +
      \mu_0\,\vec{j}(\vec{r},t) 
   \end{align}
\end{subequations}
where $\vec{E}(\vec{r},t)$ and $\vec{B}(\vec{r},t)$ denote the electric field
and the magnetic flux density, respectively. For the present case of a current
carrying lightning channel parametrised by its length coordinate, the current
density $\vec{j}(\vec{r},t)$ can be expressed as 
\begin{equation}
   \vec{j}(\vec{r},t)\,\mathrm{d}^3r = i(s,t)\,\mathrm{d}\vec{l}(s) = i(s,t)\frac{\partial\vec{l}(s)}{\partial s}\,\mathrm{d}s
\end{equation}
where $i(s,t)$ denotes the return stroke current distribution. There is no
need to replace also the charge density $\rho(\vec{r},t)$. This is due to the
fact that the scalar potential $\varPhi(\vec{r},t)$ introduced by equation
\eqref{eqefield} can directly be calculated from the Lorentz gauge
(cf. eq. \eqref{eqgauge}). Equations \eqref{eqmaxwell} are solved by the
introduction of the potentials $\varPhi(\vec{r},t)$ and $\vec{A}(\vec{r},t)$
defined through 
\begin{subequations}
   \label{eqpotential}
   \begin{align}
      \label{eqefield}
      &\vec{E}(\vec{r},t) = -\vec{\nabla}\varPhi(\vec{r},t) - \frac{\partial \vec{A}(\vec{r},t)}{\partial t}\\
      &\vec{B}(\vec{r},t) = \vec{\nabla} \times \vec{A}(\vec{r},t)
   \end{align}
\end{subequations}
Using the Lorentz gauge
\begin{equation}
\label{eqgauge}
   \vec{\nabla}\cdot\vec{A}(\vec{r},t) = -\frac{1}{c^2}\frac{\partial \varPhi(\vec{r},t)}{\partial t}
\end{equation}
and combining equations \eqref{eqpotential} and \eqref{eqmaxwell},
$\varPhi(\vec{r},t)$ and $\vec{A}(\vec{r},t)$ can be expressed as 
\begin{subequations}
   \label{eqpotential1}
   \begin{align}
      \varPhi(\vec{r},t) &= -\frac{1}{4\pi\varepsilon_0}\int\limits_0^{L_\mathrm{ret}(t)}\int\limits_{t_\mathrm{b}}^t\mathrm{d}s\:\mathrm{d}\tau \left[ \vec{\nabla} \cdot \frac{i(s,t_\mathrm{ret}(s,\tau))}{R(s)}\,\frac{\partial\vec{l}(s)}{\partial s}\right]\\
      \vec{A}(\vec{r},t) &=
      \frac{\mu_0}{4\pi}\int\limits_0^{L_\mathrm{ret}(t)} \mathrm{d}s \left[
        \frac{i(s,t_\mathrm{ret}(s,t))}{R(s)}\,\frac{\partial\vec{l}(s)}{\partial
          s} \right] 
   \end{align}
\end{subequations}
where $t_\mathrm{b}=\left|\vec{r}\right|/c$ denotes the time taken by a signal
to propagate from the lightning striking point to the observation point
$\vec{r}$, and $t_\mathrm{ret}=t-R(s)/c$ is the retarded time. Since
retardation effects have to be taken into account, the $s$-integration is
carried out along the retarded channel, that is, the range of integration is
given by $[s=0,s=L_\mathrm{ret}(t)]$. Combining equations \eqref{eqpotential}
and \eqref{eqpotential1} yields 
\begin{subequations}
\label{eqfield}
\begin{equation}
\label{eqelectricfield}
   \begin{split}
       E_k&(\vec{r},t)
       = \frac{1}{4\pi\varepsilon_0}
       \int\limits_0^{L_\mathrm{ret}(t)}\mathrm{d}s \, \Bigg\{ \\ \phantom{+[}
         \Bigg[ &\sum\limits_{j=x,y,z} \left( \frac{(r_j - l_j)(r_k -
             l_k)}{c^2 R^3} \frac{\partial l_j}{\partial s} \right) -
           \frac{1}{c^2 R} \frac{\partial l_k}{\partial s} \Bigg]
         \frac{\mathrm{d}}{\mathrm{d}t} i(s,t_{\mathrm{ret}}) \\ 
       \phantom{\Bigg[} + &\sum\limits_{j=x,y,z} \left( \frac{3(r_j - l_j)(r_k
           - l_k) - R^2\, \delta_{jk}}{c R^4} \frac{\partial l_j}{\partial s}
         \right) i(s,t_{\mathrm{ret}}) \\ 
       \phantom{\Bigg[} + &\sum\limits_{j=x,y,z} \left( \frac{3(r_j - l_j)(r_k
           - l_k) - R^2\, \delta_{jk}}{R^5} \frac{\partial l_j}{\partial s}
         \right) \int\limits_{t_b}^{t} {\mathrm{d}\tau \,
           i(s,t_{\mathrm{ret}}} \Bigg\} 
   \end{split}
\end{equation}
\begin{equation}
\label{eqmagneticfield}
   \begin{split}
        B_k(\vec{r},t)
	= \frac{\mu_0}{4\pi} \int\limits_0^{L_\mathrm{ret}(t)}\mathrm{d}s
        \Bigg[ & \left( \frac{1}{R^3}i(s,t_{\mathrm{ret}}) + \frac{1}{c
            R^2}\frac{\mathrm{d}}{\mathrm{d}t}i(s,t_{\mathrm{ret}})\right)\\&
          \sum\limits_{l,m=x,y,z}(l_l - r_l) \frac{\partial l_m}{\partial s}
          \varepsilon_{klm} \Bigg] 
    \end{split}
\end{equation}
\end{subequations}
where the Levi-Civita symbol $\varepsilon_{klm}$ is given by
\begin{equation}
   \varepsilon_{klm} = \begin{cases}	\phantom{-}	1&
     \mathrm{if}\quad(k,l,m)=(x,y,z),(y,z,x),(z,x,y)\\ -1&
     \mathrm{if}\quad(k,l,m)=(z,y,x),(y,x,z),(x,z,y)\\	\phantom{-}	0&
     \mathrm{otherwise} \end{cases} 
\end{equation}
For the sake of a better oversight the $\vec{r}$-, $s$-, and $t$-dependences
of $\vec{l}(s)$, $R(s)$, and $t_\mathrm{ret}(\vec{r},t)$ are omitted in
equations \eqref{eqfield}. 

Details to the presented solution method of Maxwell's equations can be found in many textbooks \citep[e.g.][]{wachter2005}.

Similar equations for lightning electromagnetic fields are presented in
several publications, most of them based on the assumptions of a straight and
vertical lightning channel and an observation point on ground (i.e. $z=0$)
\citep[e.g.][]{thottappillil1997,thottappillil1998,thottappillil2001,thottappillil2004,thottappillil2007}.  

Zhao and Zhang~\cite{zhao2009} derived equations for the electromagnetic fields radiated by
arbitrarily shaped lightning channels. Their approach is very similar to that
used in our work. In contrast to our work, however, their equations do not
consider arbitrarily located observation points.  

Equations \eqref{eqfield} are thus the most general expressions for lightning
return stroke fields and include all other formulations as special
cases. Assuming a straight and vertical lightning channel
(i.e. $\vec{l}(s)=s\,\vec{\mathrm{e}}_z$) and an observation point on ground
(i.e. $z=0$), for example, equations \eqref{eqfield} pass into those usually
presented in the literature
\citep[e.g.][]{thottappillil1997,thottappillil1998,rakov1998}. On the other
hand, equations \eqref{eqfield} correspond to the equations presented by
Zhao and Zhang~\cite{zhao2009} if an observation point on ground (i.e. $z=0$) is assumed.  

It should be mentioned that equations \eqref{eqfield} do not include effects
arising from a conducting ground plane (cf. subsec. \ref{conductingground}).

\subsection{The presence of a conducting ground plane}
\label{conductingground}

The presence of earth's surface is not yet included in equations
\eqref{eqfield}. Usually, the ground plane is taken into account by
considering an image channel carrying a current of opposite polarity and
direction \citep[e.g.][]{thottappillil2004}. The fields at a given observation
point then consist of the contributions from the real and the image
channel. For an observer on a perfectly conducting ground plane the
contributions of real and image channel are of equal magnitude, so that the
vertical electric field and the horizontal magnetic flux density are given by
the double of $E_z$, $B_x$, and $B_y$ from equations \eqref{eqfield}. The
horizontal electric field and the vertical magnetic flux density vanish on the
surface of a perfect conductor. When considering observation points above
ground, the contributions of real and image channel have to be calculated
separately.

\subsection{Real and retarded channel lengths for an extending lightning channel}

In order to calculate the real and retarded channel lengths for a return
stroke channel extending with velocity $v(s)$, the average velocity of the
return stroke front between ground and length coordinate $s$ is introduced
\citep{thottappillil1994} 
\begin{equation}
   v_\mathrm{av}(s) = \frac{s}{t_\mathrm{u}(s)} = \frac{s}{\int\limits_0^s\frac{\mathrm{d}\sigma}{v(\sigma)}}
\end{equation}
$t_\mathrm{u}(s)$ denotes the time taken by the return stroke front to
propagate from ground to $s$. The time depending real length $L(t)$ can be
calculated by solving the equation 
\begin{equation}
   t_\mathrm{u}(L) = \int\limits_0^{s=L}\frac{\mathrm{d}\sigma}{v(\sigma)} = t
\end{equation}
Considering retardation effects, that is, taking into account that a given
time $t$ comprises the time taken by the return stroke front to reach the
retarded length $L_\mathrm{ret}(t)$ and the time taken by the signal to
propagate to the observation point, the retarded channel length can be
calculated from \citep[cf.][]{thottappillil1994} 
\begin{equation}
  t = \frac{L_\mathrm{ret}(t)}{v_\mathrm{av}(L_\mathrm{ret}(t))}+\frac{R(L_\mathrm{ret}(t))}{c}
\end{equation}
For observation points above ground the same calculation must be done for the image channel, too.

\section{Computation of the close electromagnetic fields radiated by straight and tortuous return stroke channels}
\label{application}

\subsection{Return stroke model}

Equations \eqref{eqfield} were evaluated for a current generation type model
of the return stroke. The model of Cooray et al. \cite{cooray2004}, which refers to first
return strokes and includes the attachment process, that is, the preceding
upward growing connecting leader, was modified in order to describe subsequent
return strokes, too. For this purpose, the channel base current used by
 Cooray et al. \cite{cooray2004} was replaced by a current waveform typical for subsequent
return strokes. Moreover, the length of the upward growing connecting leader,
which  Cooray et al. \cite{cooray2004} calculated by estimating the potential difference in
the gap between connecting and dart leader, was replaced by a typical
experimental value. In simple terms, the required model input parameters are
the channel base current, the leader charge distribution, and the velocity
distribution of both connecting leader and subsequent return stroke. Details
can be found in  Cooray et al. \cite{cooray2004}. 
 
Within the framework of our field computations we investigated also the
electromagnetic fields radiated by first return strokes.  Cooray et al. \cite{cooray2004}
showed that the close electromagnetic fields associated with first return
strokes strongly depend on the assumed velocity distribution of the return
stroke front. The influence of the velocity distribution on the
electromagnetic fields radiated by subsequent return strokes was not
explicitly investigated in our work. From a physical point of view, however,
the upward growing connecting leader can be assumed to have a considerable
influence on the close electromagnetic fields, even though the connecting
leader is comparatively unpronounced for subsequent return strokes. 

\begin{figure}
   \includegraphics[width=6.5cm,height=6.5cm]{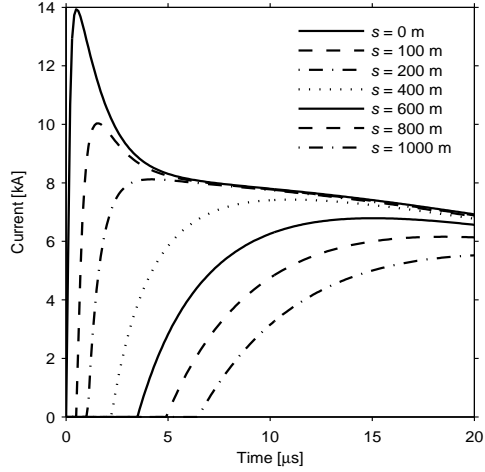}
   \caption{\it Return stroke current at different length coordinates in the
     lightning channel. The waveform associated with $s=0\,\mathrm{m}$
     corresponds to the assumed channel base current.}   
   \label{fig2}
\end{figure}

\begin{figure}
   \includegraphics[width=6.5cm,height=6.5cm]{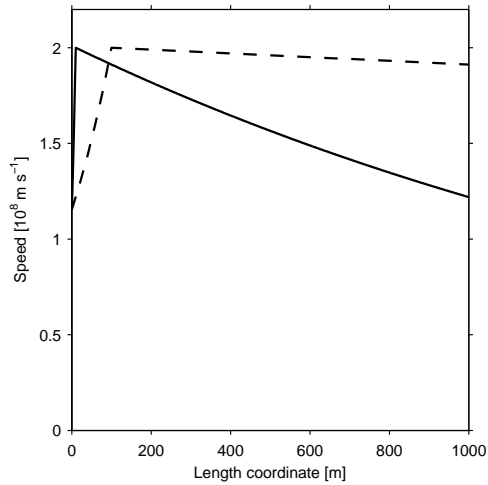}
   \caption{\it Return stroke velocity distribution as a function of the length
     coordinate $s$. The dashed line shows the velocity on a ten times faster
     length scale where the portion of the upward growing connecting leader
     becomes visible.}   
   \label{fig3}
\end{figure}

The channel base current used in our computations is shown in figure
\ref{fig2}. It consists of the sum of two Heidler functions
\citep[cf.][]{rakov1998} with a parameter set introduced by Diendorfer and Uman
\cite{diendorfer1990}. The leader charge distribution was assumed to take the
constant value of $0.14\,\mathrm{mC}\,\mathrm{m}^{-1}$. The length of the
connecting leader was set to 10\,m. These experimental values were taken from
Rakov and Uman \cite{rakov2006}. The calculated velocity distribution is shown in figure
\ref{fig3}. 

The return stroke model was implemented in Matlab 7.6 (The MathWorks Inc.,
Natick, US-MA). The analytical formulation and the numerical methods in
solving the appearing integral equation for the channel current were adapted
from Thottappillil and Uman \cite{thottappillil1994} and Cooray \cite{cooray1998}. The computed return
stroke current distribution is shown in figure \ref{fig2}. 

\subsection{Tortuous lightning channel}

\begin{figure}
   \includegraphics[width=2.7cm,height=8.0cm]{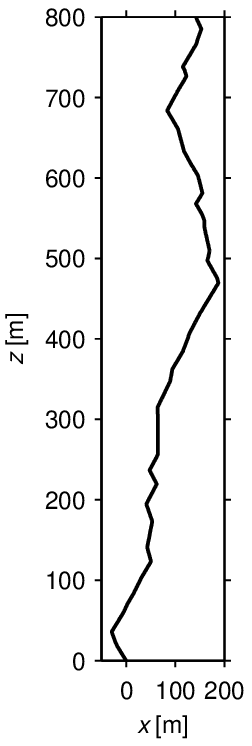}\hspace{2.0cm}
   \includegraphics[width=2.7cm,height=8.0cm]{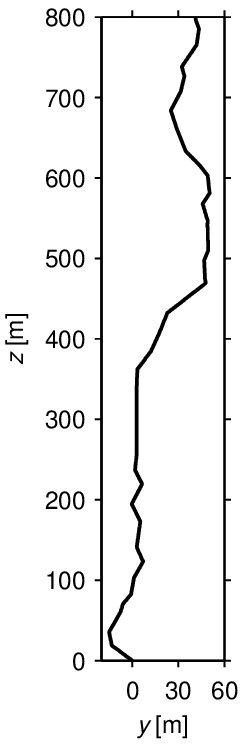}\hspace{2.0cm}
   \includegraphics[width=2.7cm,height=8.0cm]{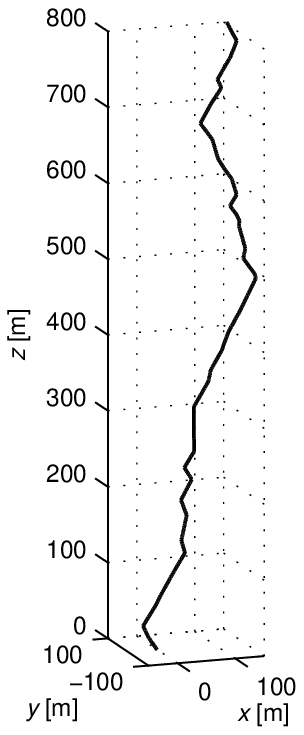}
   \caption{\it 2D and 3D perspectives of the tortuous lightning channel used for
     the field computations. The tortuosity was randomly generated. Note that
     the lowest channel sections (roughly the first 50\,m) are inclined away
     from an observer on the positive $x$-axis and inclined toward an observer
     on the negative $x$-axis.}   
   \label{fig4}
\end{figure}

The tortuous lightning channel used in our field computations was formed by a
series of straight segments. The lengths of the single segments were generated
randomly from a Gaussian distribution with a mean value of 20\,m and a
standard deviation of 10\,m. The inclination angle of each segment, that is,
the angle between the segment and the $z$-axis, was taken from a Gaussian
distribution with a mean value of 16$^\circ$  and a standard deviation of
11.5$^\circ$. The azimuth was assumed to be uniformly distributed between
0$^\circ$  and 360$^\circ$ . These parameters were chosen on the basis of
those used by Lupo et al.~\cite{lupo2000} and Vargas and Torres \cite{vargas2008}. The resulting tortuous
channel is shown in figure \ref{fig4}. 

\subsection{Radiated electromagnetic fields}
\label{radiatedfields}

Return stroke electromagnetic fields were computed for both a straight and
vertical, and a tortuous channel. In the following discussion the indices
``tort'' and ``str'' refer to the fields associated with the tortuous and the
straight lightning channel, respectively.  

The results for the electric fields and the magnetic flux densities are shown
in figures \ref{fig5} and \ref{fig6}. The observation points were assumed to
be on ground along the $x$-axis, that is, $\lvert x\rvert$ gives the distance
from the lightning striking point. While for the straight and vertical
lightning channel the magnetic flux density in $x$-direction vanishes due to
the radial symmetry, the tortuous channel radiates small magnetic fields in
$x$-direction, too. Due to the assumption of a perfectly conducting ground
plane, the horizontal electric fields vanish for both channels. 

According to figures \ref{fig5} and \ref{fig6}, the fields associated with the
straight and the tortuous channel mainly differ in the maximum
amplitudes. Moreover, the fields radiated by the tortuous channel show an
additional fine structure which changes the waveforms of the radiated
electromagnetic fields. In the following paragraphs these properties are
discussed in more detail.  

Zhao and Zhang \cite{zhao2009} presented a similar discussion of the electromagnetic fields
radiated by tortuous return stroke channels. They investigated the azimuthal
dependency of the remote fields ($100\,\mathrm{m}-100\,\mathrm{km}$ from the
lightning striking point). Our work, on the contrary, concentrates on the
radial dependency of the very close return stroke fields ($20-100\,\mathrm{m}$
from the lightning striking point). Thus, it can be regarded as a supplement
to the discussion of Zhao and Zhang \cite{zhao2009}.

\subsubsection{Electric fields}
\label{electricfields}

\begin{figure}
   \subfigure[]{\label{fig5a}\includegraphics[width=5.4cm,height=5.4cm]{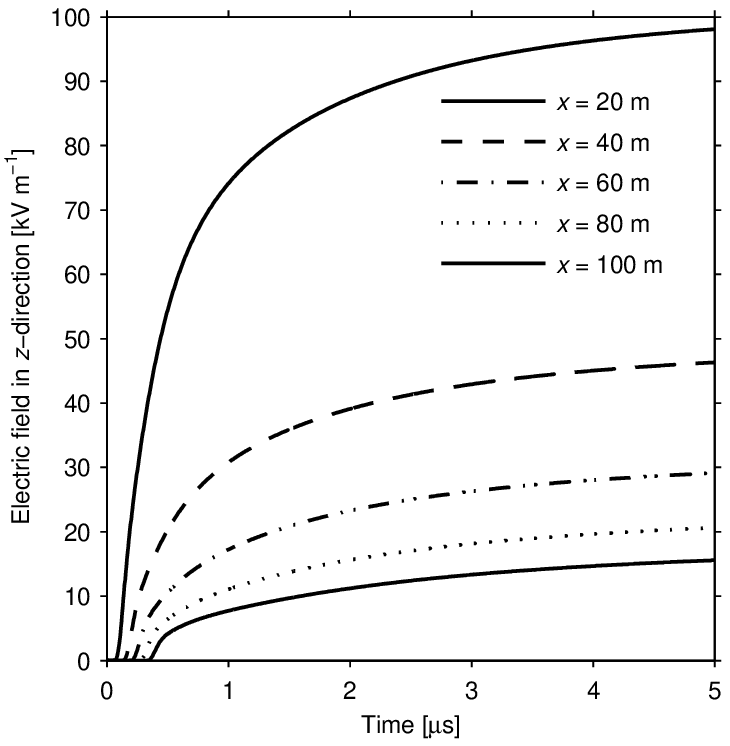}}
   \subfigure[]{\label{fig5b}\includegraphics[width=5.4cm,height=5.4cm]{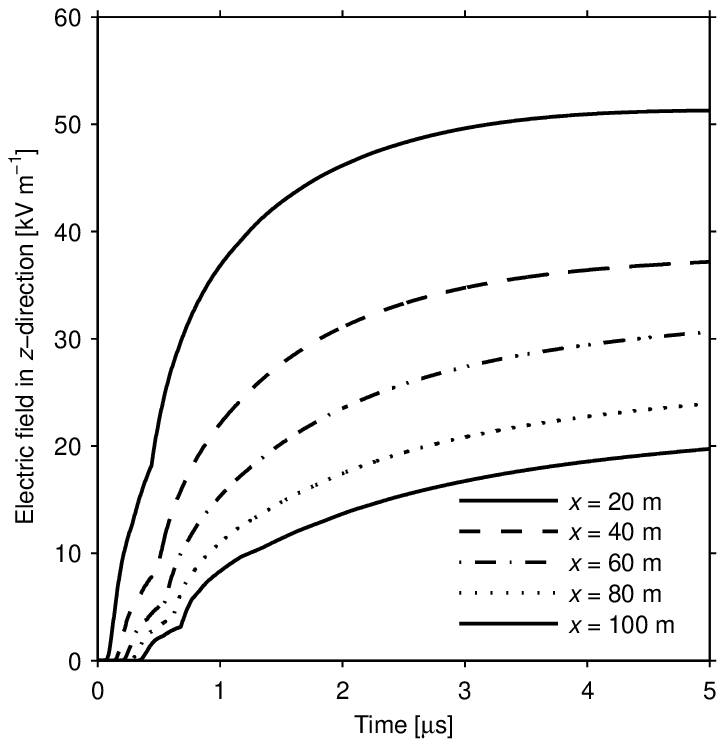}}
   \subfigure[]{\label{fig5c}\includegraphics[width=5.4cm,height=5.4cm]{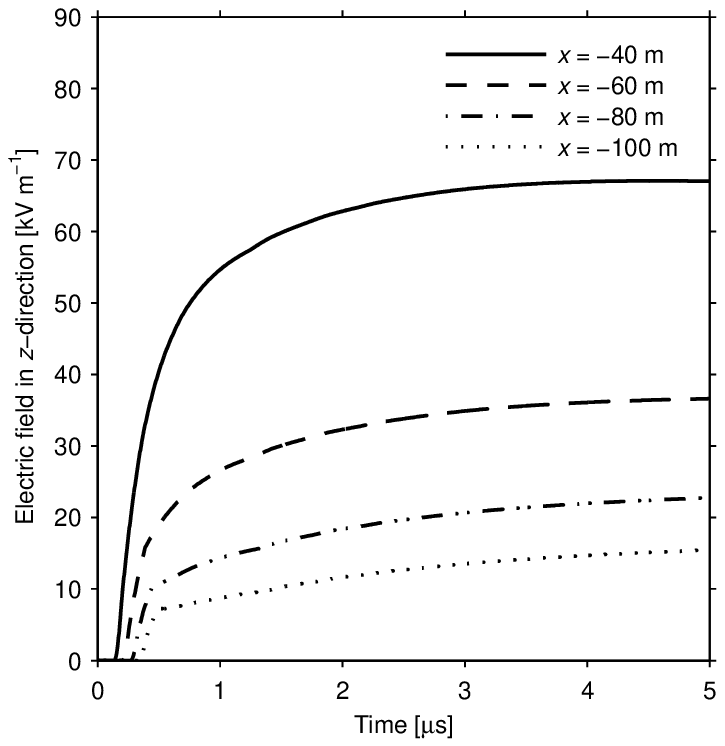}}
   \caption{\it Computed return stroke electric fields for a straight (a) and a
     tortuous (b) lightning channel. The observer was assumed to be located on
     different positions along the $x$-axis. Due to the assumption of a
     perfectly conducting ground plane, the horizontal electric fields
     vanish.} 
   \label{fig5}
\end{figure}

In the limit $t\rightarrow\infty$ the close electric fields shown in figure
\ref{fig5} converge to a constant value. This means that they are mainly
electrostatic, that is, they are dominated by the electrostatic part of
equation \eqref{eqelectricfield} (i.e. the fourth line in
eq. \eqref{eqelectricfield}), which is the only part providing a finite
contribution for $t\rightarrow\infty$. Note that the electrostatic part is
proportional to $R^{-5}$, while the other parts of equation
\eqref{eqelectricfield} are proportional to $R^{-4}$ and $R^{-3}$,
respectively. Thus, the electrostatic part mainly contributes to the close
electric fields. 

The maximum amplitudes of the radiated electric fields considerably depend on
the observation point (cf. fig. \ref{fig5}). At $x=20\,\mathrm{m}$, for
example, the maximum amplitude of $E_\mathrm{tort}$ reaches only $52\,\%$ of
the value associated with $E_\mathrm{str}$. At $x=-20\,\mathrm{m}$, on the
other hand, the maximum amplitude of $E_\mathrm{tort}$ is 1.xxx times larger
than that of $E_\mathrm{str}$. Both, the increase of $E_\mathrm{tort}$ for
observation points along the positive $x$-axis and the decrease of
$E_\mathrm{tort}$ for observation points along the negative $x$-axis are due
to the shape of the tortuous lightning channel: The lowest sections (i.e. the
first 50\,m) of the sample channel shown in figure \ref{fig4} are inclined
toward an observation point along the positive $x$-axis and inclined away from
an observation point on the negative $x$-axis. After the first 50\,m the
channel inclination is reversed. This is reflected in the electric fields at
$x=100\,\mathrm{m}$, where the $E_\mathrm{tort}$ is larger than
$E_\mathrm{str}$. The close electromagnetic fields radiated by tortuous
lightning channels thus strongly depend on the orientation of those channel
segments which primarily contribute to the total fields a given observation
point.

\subsubsection{Magnetic fields}
\label{magneticfields}

\begin{figure}
   \subfigure[]{\label{fig6a}\includegraphics[width=5.4cm,height=5.4cm]{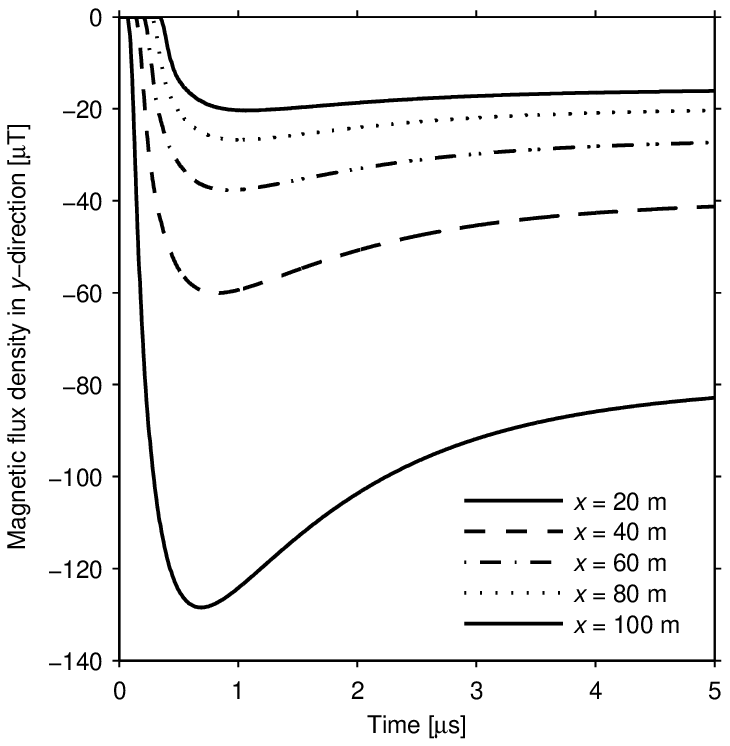}}
   \subfigure[]{\label{fig6b}\includegraphics[width=5.4cm,height=5.4cm]{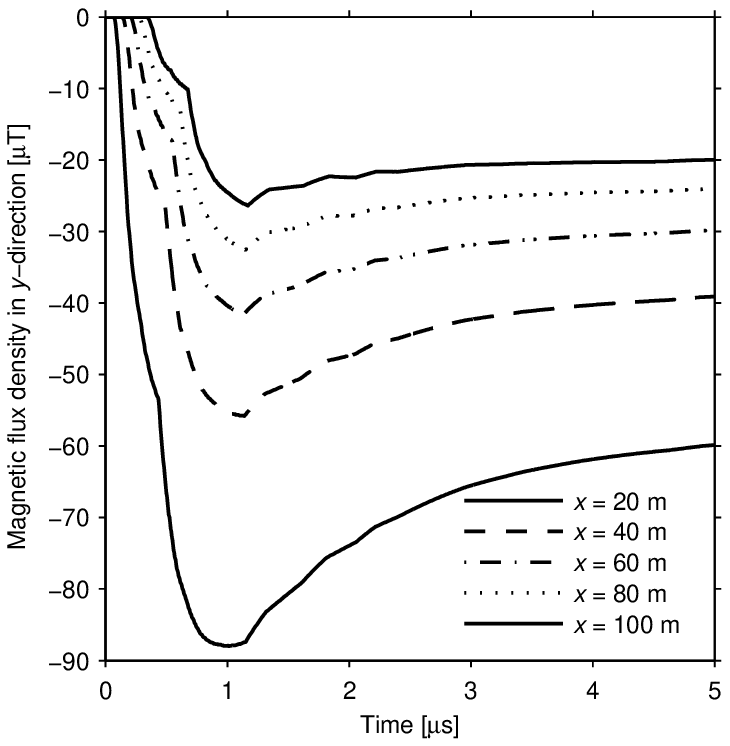}}
   \subfigure[]{\label{fig6c}\includegraphics[width=5.4cm,height=5.4cm]{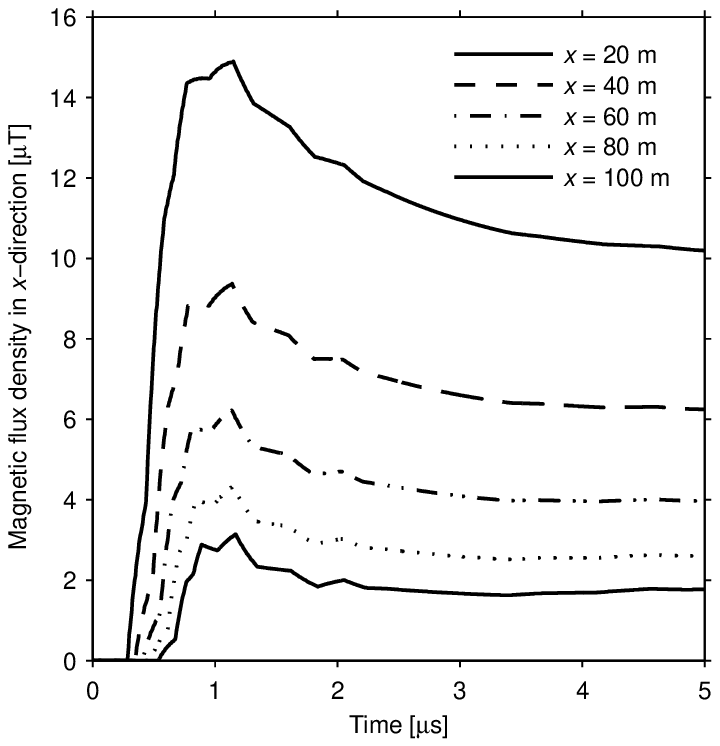}}
   \subfigure[]{\label{fig6d}\includegraphics[width=5.4cm,height=5.4cm]{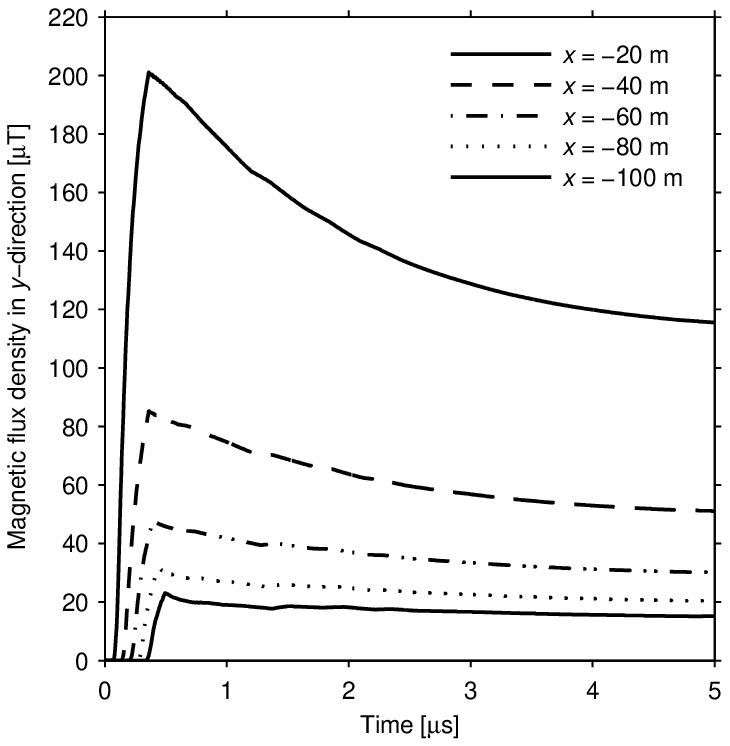}}
   \subfigure[]{\label{fig6e}\includegraphics[width=5.4cm,height=5.4cm]{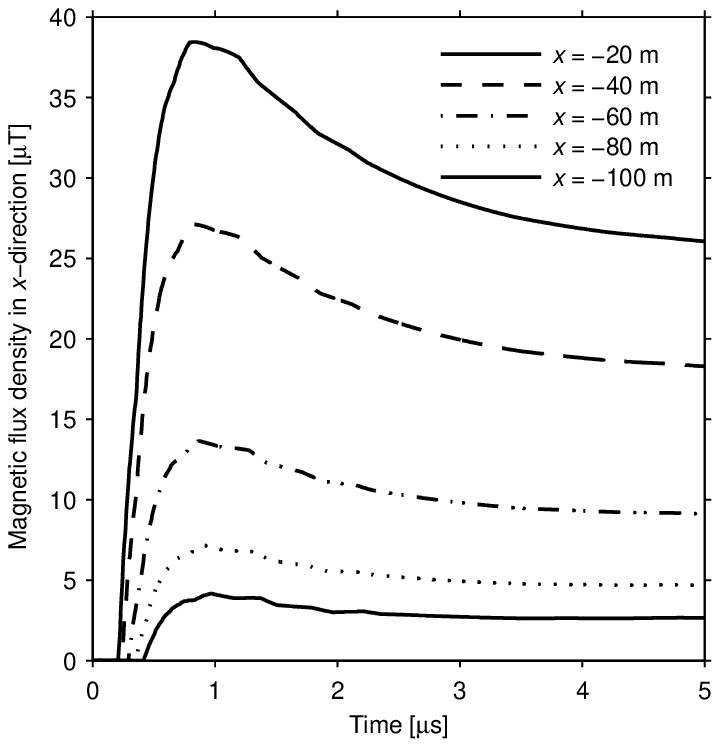}}
   \caption{\it Computed return stroke magnetic flux densities for a straight (a)
     and a tortuous (b, c, d, e) lightning channel. The observer was assumed
     to be located on different positions along the $x$-axis. The vertical
     magnetic flux densities of real and image channel cancel on a perfectly
     conducting ground plane.} 
   \label{fig6}
\end{figure}

The orientation of the lowest channel segments is reflected in the magnetic
fields, too (cf. fig. \ref{fig6}). At $x=20\,\mathrm{m}$, the maximum value of
$B_\mathrm{tort}$ reaches roughly 70\,\% of the value associated with
$B_\mathrm{str}$. At $x=20\,\mathrm{m}$, on the other hand, the peak value of
$B_\mathrm{tort}$ is more than 1.5 times larger than the peak value of
$B_\mathrm{str}$. Since the assumed tortuous channel is ``quite'' straight,
the main portion ($\sim85\,\%$) of the magnetic fields is radiated in
$y$-direction (cf. fig. \ref{fig6b}, fig. \ref{fig6c}, and fig. \ref{fig6d},
fig. \ref{fig6e}). 

The magnetic fields associated with the tortuous channel show a pronounced
fine structure. The relative amplitudes of the additional maxima are observed
to increase with increasing distance from the lightning striking point (most
clearly visible in fig. \ref{fig6c}). This simply reflects the fact that the
very close magnetic fields are mainly dominated by the lowest channel
segments, while with increasing distance also the higher channel sections
contribute. 

The peak values of $B_\mathrm{tort}$ are furthermore observed to be shifted
with respect to the peak values of $B_\mathrm{str}$ (cf. fig. \ref{fig6a},
fig. \ref{fig6b}, fig. \ref{fig6d}). At $x=20\,\mathrm{m}$ the peak value of
$B_\mathrm{tort}$ is reached $0.3\,\mu\mathrm{s}$ later than that of
$B_\mathrm{str}$. At $x=-20\,\mathrm{m}$ it is reached $0.3\,\mu\mathrm{s}$
earlier. The magnitude of this shift is observed to decrease with increasing
distance. This suggests the shift to be due to the orientation of those
channel sections which mainly contribute to the fields at a given observation
point. Generally, a shift of the peak values implicates that $B_\mathrm{tort}$
increases faster or more slowly than $B_\mathrm{str}$. If the channel sections
contributing to the fields at a given observation point are inclined away from
the observer, the fields are characterised by a more flat increase. Channel
sections inclined towards the observer, on the other hand, result in a steeper
increase. The channel orientation thus considerably effects the time
derivative of the magnetic fields. This in turn has a strong influence on the
induction component of the electric fields, that is, the electric fields
induced by the transient magnetic pulses. The effect of channel tortuosity on
the induced electric fields is discussed in subsection \ref{inducedfields}. 

\subsubsection{Induced electric fields}
\label{inducedfields}

The results of the field computations presented in the paragraphs
\ref{electricfields} and \ref{magneticfields} were used for an investigation
on the action of return stroke magnetic pulses on biological tissue. Without
going into detail it should be mentioned that the action of magnetic fields on
biological tissue mainly happens by the electric fields which are induced by
transient magnetic pulses. This paragraph presents a short discussion on the
effects of channel tortuosity on the induced electric fields. 

\begin{figure}
   \subfigure[]{\label{fig7a}\includegraphics[width=5.4cm,height=5.4cm]{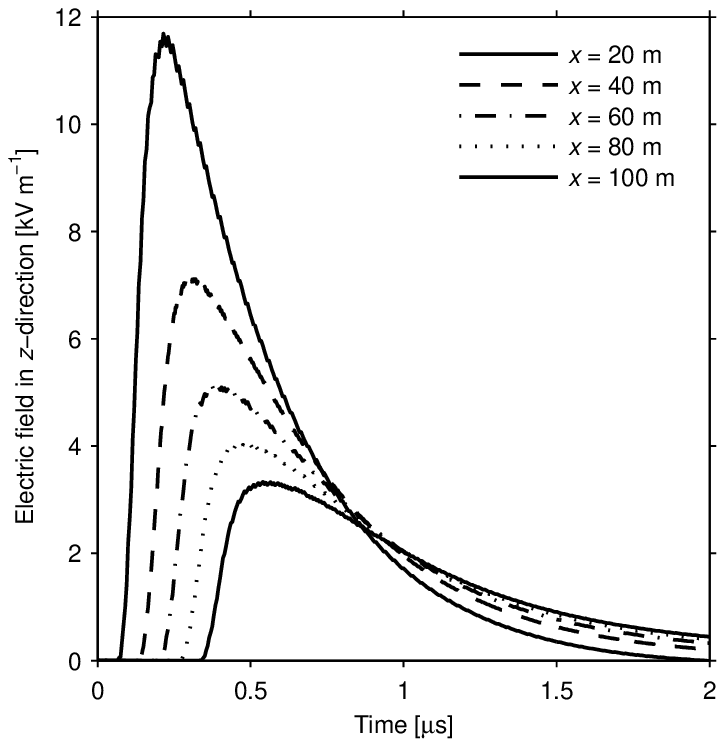}}
   \subfigure[]{\label{fig7b}\includegraphics[width=5.4cm,height=5.4cm]{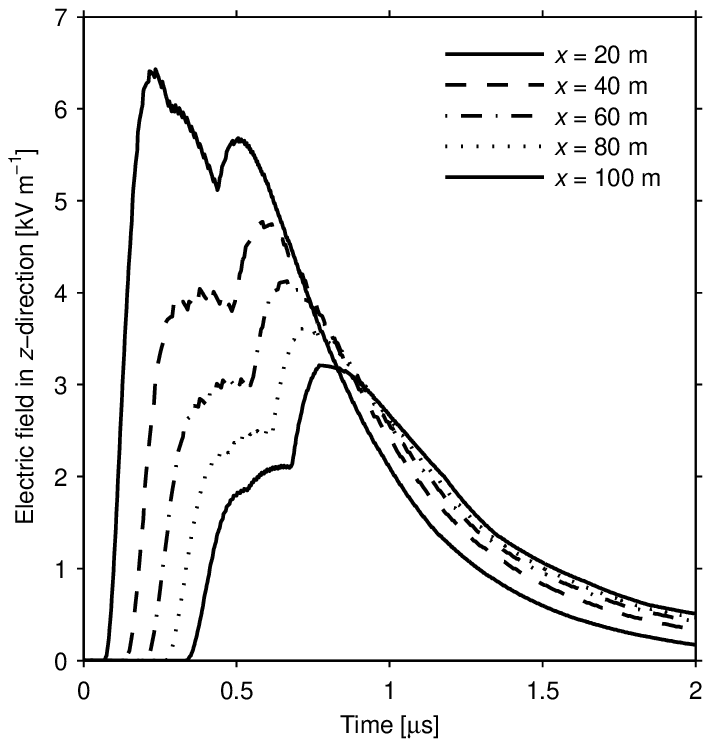}}
   \subfigure[]{\label{fig7c}\includegraphics[width=5.4cm,height=5.4cm]{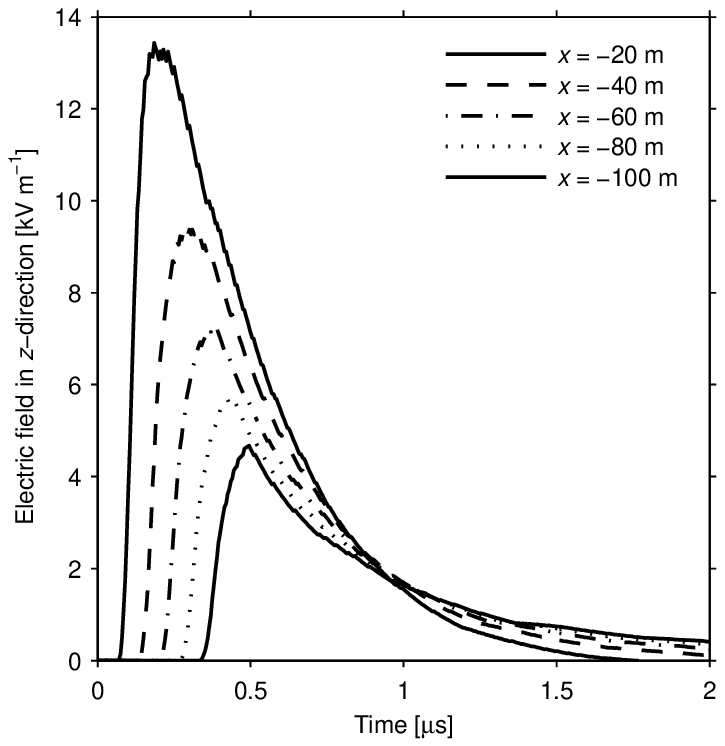}}
   \caption{\it Computed induced return stroke electric fields for a straight (a) and a tortuous (b, c) lightning channel. The observer was assumed to be located on different positions along the $x$-axis. Due to the assumption of a perfectly conducting ground plane, the horizontal induced electric fields vanish.}
   \label{fig7}
\end{figure}

The induced electric fields, in the following labelled with $E^\mathrm{ind}$,
are given by the first term of equation \eqref{eqefield}. Figure \ref{fig7}
shows the results of a numerical evaluation of this term. A comparison of
figures \ref{fig7a}, \ref{fig7b}, and \ref{fig7c} makes clear that the shape
of the lightning channel strongly influences the induced electric fields. At
$x=20\,\mathrm{m}$ the peak value of $E^\mathrm{ind}_\mathrm{tort}$ reaches
only 55\,\% of the corresponding value of $E^\mathrm{ind}_\mathrm{str}$. At
$x=-20\,\mathrm{m}$, on the other hand, the peak value of
$E^\mathrm{ind}_\mathrm{tort}$ is more than 1.1\,\% higher than that of
$E^\mathrm{ind}_\mathrm{str}$. It is furthermore observed that the fields
associated with straight and tortuous channels mainly differ in the peak
range, while their shapes are similar in the initial fast increase and the
final slow decrease. 

The induced electric fields reflect the properties of the magnetic fields
discussed in subsection \ref{magneticfields}. The magnetic fields associated
with the tortuous channel were found to increase faster or more slowly than
those associated with the straight channel (remember the shifts of the maxima
of $B_\mathrm{tort}$). Since the amplitudes of the induced electric fields
depend on the time derivative of the magnetic fields
(cf. eq. \eqref{eqinduction}), this explains the fact that the induced
electric fields are larger for observation points along the positive $x$-axis,
and smaller for observation points along the negative $x$-axis. The main
changes in the induced electric fields are due to the fastest changes in the
magnetic fields. Therefore, the main changes of the induced electric fields
concentrate on the first two microseconds. The peak value in the induced
electric fields corresponds to the point in time with the largest changes in
the magnetic fields (cf. fig. \ref{fig6}, fig. \ref{fig7}). 

For the sake of completeness it should be mentioned that the induced electric
fields are of course included in the total electric fields shown in figure
\ref{fig5}. In the close lightning environment, however, they are superimposed
by the electrostatic component and thus not explicitly visible. 

\subsection{Computational methods}

Equations \eqref{eqfield} were implemented in Matlab 7.6 (The MathWorks Inc.,
Natick, US-MA). In carrying out the numerical integrations the trapezoidal
rule was used. For the field integrations a total channel length $S$
(cf. fig. \ref{fig1}) of 1000\,m was assumed, which is enough when the fields
on ground within 100\,m from the lightning striking point are considered. Time
and length steps were set to $5\cdot10^{-9}$\,s and 1\,m, respectively. For
the fields at distances of 20\,m and 40\,m partly also higher resolutions were
needed.

\section{Summary and conclusions}
\label{conclusions}

Very general equations for lightning electric and magnetic fields including
both arbitrarily shaped lightning channels and arbitrarily located observation
points were derived from Maxwell's equations. The lightning channel was
described by a parametric representation in Cartesian coordinates. Equations
for the determination of real and retarded channel lengths for the case of a
height variable return stroke speed were presented. In order to discuss the
effect of channel tortuosity on the close return stroke electromagnetic
fields, the derived equations were solved for a current generation type model
of the lightning return stroke. Return stroke electromagnetic fields were
computed for both a straight and vertical, and a randomly generated tortuous
lightning channel. Moreover, the electric fields induced by the transient
return stroke magnetic pulses were investigated.  
 
The results show that the global amplitudes of the close electromagnetic
fields radiated by a tortuous lightning channel mainly depend on the
orientation of the lowest channel sections. The relative amplitudes of the
additional fine structure due to the channel tortuosity are observed to
increase with increasing distance from the striking point, which was reduced
to the fact that with increasing distance more different orientated channel
sections contribute to the fields at a given observation point. Furthermore,
the maximum of the magnetic fields radiated by tortuous lightning channels was
found to be shifted with respect to the magnetic fields associated with a
straight channel. The shift implicates a change in the time derivative of the
magnetic fields radiated by tortuous channels. This in turn was found to have
a remarkable influence on amplitudes and waveforms of the induced electric
fields. 



\bibliographystyle{plain}

\bibliography{lightning}

\end{document}